\documentclass[10pt]{article}
\usepackage{latexsym}
\usepackage{amssymb}
\usepackage{amsmath}
\usepackage{amscd}
\usepackage{amsthm}
\usepackage[left=2cm,top=2.5cm,right=2.5cm,bottom=1.5cm]{geometry}
\usepackage[dvips]{graphicx}
\usepackage{hyperref}

\begin{document}
\begin{center}
\Large{\bf{Symmetry Group Analysis for perfect fluid Inhomogeneous Cosmological Models in General Relativity}}

\vspace{10mm}

\normalsize{Ahmad T Ali$^{\dag, \$}$ and Anil Kumar Yadav$^{\ddag}$ \footnote{corresponding author}} \\

\vspace{4mm}
\normalsize{$^\dag$ King Abdul Aziz University,\\
Faculty of Science, Department of Mathematics,\\
PO Box 80203, Jeddah, 21589, Saudi Arabia.}\\
E-mail: atali71@yahoo.com\\
\normalsize{$^\$$ Mathematics Department,\\ Faculty of Science, Al-Azhar University,\\
Nasr city, 11884, Cairo, Egypt}\\
\vspace{2mm}
\normalsize{$^\ddag$ Department of Physics, Anand Engineering College,\\
Keetham, Agra - 282 007, India.\\
E-mail: abanilyadav@yahoo.co.in}\\
\end{center}
\begin{abstract} In this paper, we have searched the existence of the similarity
solution for plane symmetric inhomogeneous cosmological models in general relativity.
The matter source consists of perfect fluid with proportionality relation between expansion scalar
and shear scalar. The isovector fields of Einstein's field equation for the models under consideration
are derived. A new class of exact solutions of Einstein's field equation have been obtained
for inhomogeneous space-time. The physical behaviors and geometric aspects of the derived models
have been discussed in detail.

\end{abstract}

\emph{PACS:} 98.80.JK, 98.80.-k.

\emph{Keywords}: Similarity solutions, Inhomogeneous cosmological model, General relativity.

\section{Introduction }

An Inhomogeneous cosmological model plays a significant role in
understanding of phenomenon like formation of galaxies during its early stage
of evolution. The choice of inhomogeneous cosmological models permit one to obtain
more general cosmological model in comparison to FRW model and Bianchi type models of universe.
The theoretical arguments supports the existence of an inhomogeneous phase of early universe
that approaches to homogeneity on later stage of evolution. The recent observation confirmed that
the present evolution of universe is spherically symmetric and the matter distribution
in the universe is on the whole isotropic and homogeneous. But close to the big bang singularity,
neither the assumption of spherically symmetric nor that of isotropy can be strictly valid. So,
we consider plane symmetric space time with inhomogeneous distribution of matter, which is less
restrictive than spherical symmetric and can provide an opportunity for the study of inhomogeneity.\\

Firstly, Taub \cite{taub1, taub2}  considered the plane-symmetric inhomogeneous cosmological model and
later on, Fienstein and Senovilla \cite{fiens1} obtained a new class of exact solution of Einstein's field equation
with big bang singularity which satisfied energy and causality condition. A large number of singularity
free inhomogeneous cosmological models have been analysed by Senovilla \cite{seno1} Ruis and Senovilla \cite{ruis1},
through a comprehensive study of inhomogeneous metric with separable function of $r$ and $t$ as
metric coefficient. Pradhan et al \cite{pradhan00, pradhan01, pradhan1, pradhan2, pradhan3}  and Yadav \cite{yadav1}
have presented plane symmetric and cylindrically symmetric inhomogeneous cosmological models in different
physical aspects. In 2008, Marra and P\"{a}\"{a}kk\"{o}nen \cite{marra1} developed the formalism 
to study exact spherically-symmetric inhomogeneous models with an arbitrary numbers of perfect fluids and later on, 
Marra et al \cite{marra2} confronted these models with data. Recently Bolejko et al \cite{bole1} 
have studied the inhomogeneous cosmological models in
the framework of general relativity. According to them, inhomogeneous cosmological models are the
exact solution of Einstein's field equation that contains at least one subclass of non static FRW solution
as a limit.

The symmetry groups are defined as the groups of continuous
transformations that leave a given family of equations invariant \cite{ali1, ali2, att1, mekh1}.
Nonlinear equations are widely used as models to describe complex physical phenomena in various fields of 
science, especially in fluid mechanics, solid state physics, plasma physics, plasma wave and general relativity. 
The investigation of the exact solutions of nonlinear partial differential equations (PDEs) plays an
important role in the study of nonlinear physical phenomena \cite{ali0, ali3, elsa1, elsa2}. 
Although the concept of symmetry transformations is well-known in the theory of
differential equations, both ordinary and partial, we owe their
first systematic treatment to Ovsiannikov \cite{ovsi1} who had observed
that the usual Lie infinitesimal invariance approach could as well
be employed in order to construct symmetry groups \cite{blum1, ibra1, olve1}.

In this paper, we apply the so-called symmetry analysis method for a particular problem in 
general relativity. The main advantage of such method is that they can be successfully applied to 
nonlinear differential equations. The similarity solutions are quite popular because they result in 
the reduction of the independent variables of the problem. In our case, the problem under investigation 
is the system of second order nonlinear PDEs. Hence, any similarity solution will transform the system of 
nonlinear PDEs into a system of ODEs.

From the discussion above, we attempted to find a new class of exact solutions for Einstein field equations. 
A plane symmetric inhomogeneous cosmological models  with a perfect fluid in Riemannian geometry, 
are introduced in section 2. In section 3 symmetry
analysis and isovector fields for Einstein field equations are
obtained. In section 4 we found new class of exact (similarity) solutions for
Einstein field equations. Section 5 is the study of some physical and geometrical properties of the obtained model.

\section{The metric and field equations}

We consider the plane-symmetric line element in in general form
\begin{equation}  \label{u21}
ds^2=-A^2\,dt^2+B^2\,dr^2+C^2\,(dx^2+dy^2),
\end{equation}
where $A$, $B$ and $C$ are functions of $r$ and $t$. The volume element of the model (\ref{u21}) is given by
\begin{equation}  \label{u22-1}
V=\sqrt{-g}=A\,B\,C^2.
\end{equation}
The four velocity of the fluid has the form
\begin{equation}  \label{u22-2}
u^i=\Big(0,0,0,\dfrac{1}{A}\Big).
\end{equation}
The energy-momentum tensor in the presence of a perfect fluid has the form
\begin{equation}  \label{u25}
T_{ij}=(\rho+p)\,u_i\,u_j+p\,g_{ij},
\end{equation}
where $\rho$ and $p$ are the energy density and the pressure of the fluid, respectively. 
In this coordinate system the Einstein's field equation
\begin{equation}  \label{u23}
G_{ij}=R_{ij}-\dfrac{1}{2}\,g_{ij}\,R=-\chi\,T_{ij}.
\end{equation}
For the line element (\ref{u21}) the field equation (\ref{u23}) can be reduced to the 
following system of non-linear partial differential equations:
\begin{equation}  \label{u210}
\begin{array}{ll}
    \dfrac{C_{rt}}{C^2}-\dfrac{A_r\,C_t}{A\,C}-\dfrac{B_t\,C_r}{B\,C}=0,
  \end{array}
\end{equation}

\begin{equation}  \label{u211}
\begin{array}{ll}
        \dfrac{A_{rr}}{A}+\dfrac{C_{rr}}{C}-\dfrac{A_r\,B_r}{A\,B}-\dfrac{A_r\,C_r}{A\,C}-\dfrac{B_r\,C_r}{B\,C}-\dfrac{C_r^2}{C^2}\\
        \\
    \,\,\,\,\,\,\,\,\,\,\,\,\,\,\,\,\,\,\,\,\,\,\,\,\,\,\,\,\,\,\,
    -\dfrac{B^2}{A^2}\Bigg(\dfrac{B_{tt}}{B}-\dfrac{C_{tt}}{C}-\dfrac{A_t\,B_t}{A\,B}+\dfrac{A_t\,C_t}{A\,C}+\dfrac{B_t\,C_t}{B\,C}-\dfrac{C_t^2}{C^2}\Bigg)=0,
  \end{array}
\end{equation}

\begin{equation}  \label{u212}
\begin{array}{ll}
        \chi\,A^2\,\rho=\dfrac{2\,B_t\,C_t}{B\,C}+\dfrac{C_t^2}{C^2}-\dfrac{A^2}{B^2}\Bigg(\dfrac{2\,C_{rr}}{C}+\dfrac{C_r^2}{C^2}-\dfrac{2\,B_r\,C_r}{B\,C}\Bigg),
  \end{array}
\end{equation}

\begin{equation}  \label{u213}
\begin{array}{ll}
        \chi\,B^2\,p=\dfrac{2\,A_r\,C_r}{A\,C}+\dfrac{C_r^2}{C^2}-\dfrac{B^2}{A^2}\Bigg(\dfrac{2\,C_{tt}}{C}+\dfrac{C_t^2}{c^2}-\dfrac{2\,A_t\,C_t}{A\,C}\Big).
  \end{array}
\end{equation}
The consequence of the energy momentum conservation
\begin{equation}
\begin{array}{ll}
T^{ij}_{;j}=0
  \end{array}
\end{equation}
are the relations
\begin{equation}
\left\{
\begin{array}{ll}
p_r+(\rho+p)\,\dfrac{A_r}{A}=0,\\
\\
\rho_t+(\rho+p)\,\Big(\dfrac{B_t}{B}+\dfrac{2\,C_t}{C}\Big)=0.
\end{array}
\right.
\end{equation}
The plane symmetric solutions can be classified according to their four kinematic properties, i.e., 
rotation, acceleration, expansion and shear. In co-moving frame of reference these quantities can be 
computed as the following:
\begin{equation}  \label{u22-3}
    \dot{u}_i\,=\,u_{i;j}\,u^j\,=\,\Big(\dfrac{A_r}{A},0,0,0\Big),
    \end{equation}

\begin{equation}  \label{u22-3-1}
  \omega_{ij}\,=\,u_{[i;j]}+\dot{u}_{[i}\,u_{j]}\,\equiv\,0,
   \end{equation}

   \begin{equation}  \label{u22-3-2}
  \Theta\,=\,u_{;i}^{i}=\dfrac{1}{A}\Big(\dfrac{B_t}{B}+\dfrac{2\,C_t}{C}\Big),
  \end{equation}

\begin{equation}  \label{u22-4}
  \begin{array}{ll}
\sigma_{ij}\,=\,u_{(i;j)}+\dot{u}_{(i}\,u_{j)}-\frac{1}{3}\,\Theta\,(g_{ij}+u_i\,u_j).
\end{array}
\end{equation}

Therefore the non-vanishing components of the shear tensor $\sigma_i^j$ and the shear scalar are
\begin{equation}  \label{u22-6}
  \begin{array}{ll}
    \sigma_3^3\,=\,\sigma_2^2\,=\,-\dfrac{1}{2}\,\sigma_1^1\,=\,\dfrac{1}{3\,A}\Big(\dfrac{C_t}{C}-\dfrac{B_t}{B}\Big),
   \end{array}
\end{equation}

\begin{equation}  \label{u22-6-1}
  \begin{array}{ll}
 \sigma^2\,=\,\frac{1}{2}\,\sigma_{ij}\,\sigma^{ij}=\dfrac{1}{3\,A^2}\Big(
 \dfrac{C_t}{C}-\dfrac{B_t}{B}\Big)^2.
   \end{array}
\end{equation}

The Einstein field equations (\ref{u210})-(\ref{u213}) constitute a system of four highly 
non-linear differential equations with five unknowns variables, $A$, $B$, $C$, $p$ and $\rho$. 
Therefore, one physically reasonable conditions amongst these parameters are required to obtain explicit solutions of 
the field equations. Let us assume that the expansion scalar $\Theta$ in the model (\ref{u21}) is proportional to the eigenvalue $\sigma_1^1$ of the shear tensor $\sigma_j^k$. Then from (\ref{u22-3-2}) and (\ref{u22-6}), we get
\begin{equation}\label{u216}
  \begin{array}{ll}
\dfrac{B_t}{B}-\dfrac{C_t}{C}=3\,\gamma\,\Big(\dfrac{C_t}{C}+\dfrac{B_t}{2\,B}\Big),
  \end{array}
\end{equation}
where $\gamma$ is a constant of proportionality. The above equation can be written in the form
\begin{equation}\label{u217}
  \begin{array}{ll}
\dfrac{C_t}{C}\,=\dfrac{1}{2}\Big(\dfrac{2-3\,\gamma}{1+3\,\gamma}\Big)\dfrac{B_t}{B}.
  \end{array}
\end{equation}
If we integrate the above equation with respect to $t$, we can get the following relation
\begin{equation}\label{u217}
  \begin{array}{ll}
C(r,t)\,=\,f(r)\,B^n(r,t),
  \end{array}
\end{equation}
where  $n=\dfrac{1}{2}\Big(\dfrac{2-3\,\gamma}{1+3\,\gamma}\Big)$ and $f(r)$ is a constant of integration which 
is an arbitrary function of $r$. If we substitute the metric function $C$ from (\ref{u217}) in 
the Einstein field equations, the equations (\ref{u210})-(\ref{u211}) transform to the nonlinear partial 
differential equations of the coefficients $A$ and $B$ only, as the following new form:
\begin{equation}  \label{u210-1}
\begin{array}{ll}
    E_1=\dfrac{B_{rt}}{B}+\dfrac{(n-2)\,B_r\,B_t}{B^2}-\dfrac{A_r\,B_t}{A\,B}+\dfrac{(n-1)\,f'\,B_t}{n\,f\,B}=0,
  \end{array}
\end{equation}

\begin{equation}  \label{u211-1}
\begin{array}{ll}
        E_2=\dfrac{(n-1)\,B^2}{A^2}\Bigg[\dfrac{B_{tt}}{n\,B}+\dfrac{2\,B_t^2}{B^2}-\dfrac{A_t\,B_t}{n\,A\,B}\Bigg]
        +\dfrac{B_{rr}}{B}+\dfrac{A_{rr}}{n\,A}-\dfrac{2\,B_r^2}{B^2}\\
        \\
        \,\,\,\,\,\,\,\,\,\,\,\,\,\,\,\,\,\,\,\,\,\,\,\,\,\,\,\,\,\,\,\,\,\,\,\,\,\,\,\,
        \,\,\,\,\,\,\,\,\,\,\,\,\,\,\,\,\,\,\,\,
       -\dfrac{(n+1)\,A_r\,B_r}{n\,A\,B}-\dfrac{f'}{n\,f}\Big(\dfrac{B_r}{B}+\dfrac{A_r}{A}\Big)+\dfrac{f''}{n\,f}-\dfrac{f'^2}{n\,f^2}=0,
  \end{array}
\end{equation}
where the prime indicates derivative with respect to the coordinate $r$.

\section{Symmetry analysis method}

In order to obtain an exact solutions of the system of nonlinear partial 
differential equations (\ref{u210-1})-(\ref{u211-1}), we will use the symmetry
analysis method. For this we write
\begin{equation}\label{u31}
\left\{
\begin{array}{ll}
x_i^{*}=x_i+\epsilon\,\xi_{i}(x_j,u_{\beta})+\bold{o}(\epsilon^2),\\
u_{\alpha}^{*}=u_{\alpha}+\epsilon\,\eta_{\alpha}(x_j,u_{\beta})+\bold{o}(\epsilon^2),
\end{array}
\right.
\,\,\,i,j,\alpha,\beta=1,2,
\end{equation}
as the infinitesimal Lie point transformations. We have assumed
that the system (\ref{u210-1})-(\ref{u211-1}) is invariant under the transformations given in
Eq. (\ref{u31}). The corresponding infinitesimal generator of Lie groups
(symmetries) is given by
\begin{equation}\label{u32}
 X=\sum_{i=1}^{2}\xi_{i}\dfrac{\partial}{\partial x_{i}}+\sum_{\alpha=1}^{2}\eta_{\alpha}
 \dfrac{\partial}{\partial u_{\alpha}},
 \end{equation}
where $x_1=r$, $x_2=t$, $u_1=A$ and $u_2=B$. The coefficients $\xi_{1}$, $\xi_{2}$, $\eta_{1}$ and $\eta_{2}$ are the functions of $r$, $t$, $A$ and $B$.
These coefficients are the components of infinitesimals symmetries
corresponding to $r$, $t$, $A$ and $B$ respectively, to be determined from the invariance conditions:
\begin{equation}\label{u33}
{\text{Pr}}^{(2)}\,X\Big(E_m\Big)|_{E_m=0}=0,
\end{equation}
where $E_m=0,\,m=1,2$ are the system (\ref{u210-1})-(\ref{u211-1}) under study and
${\text{Pr}}^{(2)}$ is the second prolongation of the symmetries $X$.
Since our equations (\ref{u210-1})-(\ref{u211-1}) are at most of order two, therefore, we
need second order prolongation of the infinitesimal generator
in Eq. (\ref{u33}). It is worth noting that, the $n$-th order prolongation is given by:
\begin{equation}\label{u34}
{\text{Pr}}^{(n)}\,X=X+\sum_{s=1}^{n}\,\sum_{\alpha=1}^{3}\,\eta_{\alpha,i_1i_2...i_s}\,\dfrac{\partial}{\partial u_{\alpha,i_1i_2...i_s}},
\end{equation}
where
\begin{equation}\label{u35}
\eta_{\alpha,i_1i_2...i_s}=D_{i_1i_2...i_s}\Big[\eta_{\alpha}-\sum_{i=1}^{2}
\,\xi_i\,u_{\alpha,i}\Big]+\sum_{i=1}^{2}\,\xi_{i}\,u_{\alpha,i_1i_2...i_si}\,.
\end{equation}
The operator $D_{i_1i_2...i_s}$ is called the {\it total derivative} ({\it Hach operator}) and taken the following
form:
\begin{equation}\label{u36}
D_i=\dfrac{\partial}{\partial x_i}+\sum_{s=1}^{n}\,\sum_{\alpha=1}^{3}\,u_{\alpha,i_1i_2...i_s}\,
\dfrac{\partial}{\partial u_{\alpha,i_1i_2...i_s}},
\end{equation}
where $D_{ij}=D_{ji}$ and $u_{\alpha,i}=\frac{\partial u_{\alpha}}{\partial x_{i}}$.

Expanding the system of Eqs. (\ref{u33}) with the original system of Eqs. (\ref{u210-1})-(\ref{u211-1}) 
to eliminate $B_{rr}$ and $B_{rt}$ while we set the 
coefficients involving $A_{r}$, $A_{t}$, $A_{rr}$, $A_{rt}$, $A_{tt}$,
$B_{r}$, $B_{t}$, $B_{tt}$ and
various products to zero give rise the essential set of over-determined
equations. Solving the set of these determining equations, the components of symmetries takes the following form:
\begin{equation}\label{u37}
\xi_{1}=c_1\,r+c_2,\,\,\,\,\,\xi_{2}=c_3\,t+c_4,\,\,\,\,\,\eta_{1}=(c_1+c_3)\,A ,\,\,\,\,\,\eta_{2}=2\,c_3\,B,
\end{equation}
such that the function $f(r)$ must be equal:
\begin{equation}\label{u37-1}\left\{
                               \begin{array}{ll}
                                 f(r)=c_5\,\exp\big[c_6\,r\big],\,\,\,\,\,\,\,\,\,\,\,\mathrm{if}\,\,\, c_1=0, \\
\\
                                 f(r)=c_7\big(c_1\,r+c_2\big)^{c_8},\,\,\,\,\,\,\mathrm{if}\,\,\, c_1\neq0,
                               \end{array}
                             \right.
\end{equation}
where $c_i,\,i=1,2,...,8$ are an arbitrary constants.

\section{Similarity solutions}

The characteristic equations corresponding to the symmetries (\ref{u37}) are given by:
\begin{equation}\label{u41-1}
\dfrac{dx}{c_1\,r+c_2}=\dfrac{dt}{c_3\,t+c_4}=\dfrac{dA}{(c_1+c_3)\,A}=\dfrac{dB}{2\,c_3\,B}.
\end{equation}
By solving the above system, we have the following four cases:\\

\textbf{Case (1):} When $c_1=c_3=0$, the similarity variable and similarity functions can be written as the following:
\begin{equation}\label{u42-1}
\begin{array}{ll}
\xi=a\,r+b\,t,\,\,\,\,\,\,A(r,t)=\Psi(\xi),\,\,\,\,\,\,B(r,t)=\Phi(\xi),
\end{array}
\end{equation}
where $a=c_4$ and $b=-c_2$ are an arbitrary constants. Substituting the transformations (\ref{u41-1}) in the field Eqs. (\ref{u210-1})-(\ref{u211-1}) lead
to the following system of ordinary differential equations:
\begin{equation}\label{u43-1}
\begin{array}{ll}
\dfrac{\Phi''}{\Phi}-\dfrac{\Phi'\,\Psi'}{\Phi\,\Psi}+(n-2)\,\Big(\dfrac{\Phi'^2}{\Phi^2}\Big)+\dfrac{d\,(n-1)}{a\,n}\,\,\Big(\dfrac{\Phi'}{\Phi}\Big)=0,
\end{array}
\end{equation}

\begin{equation}\label{u44-1}
\begin{array}{ll}
\dfrac{(1-n)\,b^2\,\Phi^2}{a^2\,\Psi^2}\Big[\dfrac{\Phi''}{\Phi}+2\,n\,\Big(\dfrac{\Phi'^2}{\Phi^2}\Big)-\dfrac{\Phi'\,\Psi'}{\Phi\,\Psi}\Big]
=\dfrac{\Psi''}{\Psi}+n\,\Big(\dfrac{\Phi''}{\Phi}\Big)\\
\\
\,\,\,\,\,\,\,\,\,\,\,\,\,\,\,\,\,\,\,\,\,\,\,\,\,\,\,\,\,\,\,\,\,\,\,\,\,\,\,\,\,\,\,\,\,\,\,\,\,\,
-(n-1)\,\Big(\dfrac{\Phi'\,\Psi'}{\Phi\,\Psi}\Big)-2\,n\,\Big(\dfrac{\Phi'^2}{\Phi^2}\Big)
-\dfrac{d}{a}\Big(\dfrac{\Phi'}{\Phi}+\dfrac{\Psi'}{\Psi}\Big),
\end{array}
\end{equation}
where $c=c_7$ and $d=c_8$ are arbitrary constants such that $f(r)=c\,\mathrm{exp}[d\,r]$. By integration the equation (\ref{u43-1}), we can get the following:
\begin{equation}\label{u45-1}
\begin{array}{ll}
\Psi(\xi)=q_1\,\Phi^{n-2}(\xi)\,\Phi'(\xi)\,\exp\big[\tilde{d}\,\xi\big],
\end{array}
\end{equation}
where $\tilde{d}=\dfrac{d\,(n-1)}{a\,n}$ while $q_1$ is an arbitrary constant of integration. Substitute (\ref{u45-1}) in (\ref{u44-1}), we have the following ordinary differential equation of the function $\Phi$ only as follows:

\begin{equation}\label{u46-1}
\begin{array}{ll}
a^2\,q_1^2\,\Phi^{2(n-3)}\,\exp\big[2\,\tilde{d}\,\xi\big]\Bigg[
2\,(n-1)\,(3\,n-4)\,\Phi'^3+(n-1)\,\Phi\,\Phi'\big[
5\,\tilde{d}\,\Phi'\\
\,\,\,\,\,\,\,\,\,\,\,\,\,\,\,\,\,\,\,\,\,\,\,\,\,\,\,\,\,\,
+(7-3\,n)\,\Phi''\big]
+\Phi^2\,\Big(
\tilde{d}^2\,\Phi'-\tilde{d}\,(n-2)\,\Phi''-(n-1)\,\Phi'''\big]
\Big)
\Bigg]\\
\,\,\,\,\,\,\,\,\,\,\,\,\,\,\,\,\,\,\,\,\,\,\,\,\,\,\,\,\,\,
=(1-n)^2\,b^2\big[(n+2)\,\Phi'-\tilde{d}\,\Phi\big].
\end{array}
\end{equation}
If one solves the above third order non-linear ordinary differential equation, he can obtain the exact solutions of the original Einstein field equations (\ref{u210})-(\ref{u213}) corresponding to reduction (\ref{u41-1}). This equation is very difficult to solve in general form, however, may be solved in some special cases.\\

Here, we assume the solution equation (\ref{u46-1}) in the form:
\begin{equation}\label{u47-1}
\begin{array}{ll}
\Phi(\xi)=q_2\,\exp[q_3\,\xi],
\end{array}
\end{equation}
where $q_2$ and $q_3$ are arbitrary non-zero constants. If we substitute the above solution in the equations (\ref{u46-1}), we have the following condition:
\begin{equation}\label{u48-1}
\begin{array}{ll}
\Big[2\,b^2\,(1-n)^2\,q_2^{2\,(2-n)}\,(\alpha_1+4\,n\,q_3)\Big]\,\exp\big[\alpha_1\,\xi\big]=a_1^2\,q_1^2\,q_3\,(2\,q_3-\alpha_1)\,(4\,n\,q_3-\alpha_1),
\end{array}
\end{equation}
where $\alpha_1=2\,\big[(2-n)\,q_3-\tilde{d}\big]$. For discuss the above condition, we must take two cases as the following:\\

\textbf{Case (1.1):} When $\alpha_1\,\neq\,0$, then the condition (\ref{u48-1}) leads to the following two conditions:
\begin{equation}\label{u49-1}
\begin{array}{ll}
\alpha_1+4\,n\,q_3=0,\,\,\,\,\,\,\,(2\,q_3-\alpha_1)\,(4\,n\,q_3-\alpha_1)=0.
\end{array}
\end{equation}
The solution of the above equation is $\alpha_1=q_3=0$ contradiction.\\

\textbf{Case (1.2):} When $\alpha_1\,=\,0$, the condition (\ref{u48-1}) leads to $q_1=\dfrac{b\,(n-1)\,q_2^{2-n}}{a\,q_3}$. Therefore, the coefficients metric functions takes the form:
\begin{equation}\label{u410-1}
\left\{
  \begin{array}{ll}
A(r,t)=a_1\,\exp[\tilde{a}\,r+\tilde{b}\,t],\\
\\
B(r,t)=a_2\,\exp[\tilde{a}\,r+\tilde{b}\,t],\\
\\
C(r,t)=a_3\,\exp\Big[\dfrac{n}{n-1}\Big(\tilde{a}\,r+\tilde{b}\,(n-1)\,t\Big)\Big],
  \end{array}
\right.
\end{equation}
where $a_1=\dfrac{q_2\,(n-1)\,\tilde{b}}{\tilde{a}}$, $a_2=q_2$ and $a_3=c_1\,q_2^{n}$.  It is observed from the above equations, the line element (\ref{u21}) can be written in the following form:
\begin{equation}  \label{s1}
\begin{array}{ll}
ds_1^2=a_2^2\,\exp\Big[2\big(\tilde{a}\,r+\tilde{b}\,t\big)\Big]\,\Big(dr^2-\dfrac{(n-1)^2\,\tilde{b}^2\,dt^2}{\tilde{a}^2}\Big)\\
\\
\,\,\,\,\,\,\,\,\,\,\,\,\,\,\,\,\,\,\,\,\,\,\,\,\,\,\,\,\,\,\,\,\,\,\,\,\,\,\,\,\,\,\,\,\,\,
\,\,\,\,\,\,\,
+a_3^2\,\exp\Big[\dfrac{2\,n}{n-1}\Big(\tilde{a}\,r+\tilde{b}\,(n-1)\,t\Big)\Big]\,\Big(dx^2+dy^2\Big).
\end{array}
\end{equation}
where $a_2$, $a_3$, $\tilde{a}$, $\tilde{b}$ and $n$ are an arbitrary constants.\\

\textbf{Case (2):} When $c_1=0$ and $c_3\neq0$, the similarity variable and similarity functions can be written as the following:
\begin{equation}\label{u51-2}
\begin{array}{ll}
\xi=(t+b)\,\exp[a\,r],\,\,\,\,\,A(r,t)=\Psi(\xi)\,\exp[-a\,r],\,\,\,\,\,B(r,t)=\Phi(\xi)\,\exp[-2\,a\,r],
\end{array}
\end{equation}
where $a=-\dfrac{c_3}{c_2}$ and $b=\dfrac{c_4}{c_3}$ are an arbitrary constants. Substituting the transformations (\ref{u51-2}) in the field Esq. (\ref{u210-1}) and (\ref{u211-1}) lead
to the following system of ordinary differential equations:
\begin{equation}\label{u52-2}
\begin{array}{ll}
\dfrac{\Phi''}{\Phi}-\dfrac{\Phi'\,\Psi'}{\Phi\,\Psi}+(n-2)\,\dfrac{\Phi'^2}{\Phi^2}
=\Big(\dfrac{(n-1)\,d+2(n-2)\,a}{a\,n\,\xi}\Big)\,\dfrac{\Phi'}{\Phi},
\end{array}
\end{equation}

\begin{equation}\label{u53-2}
\begin{array}{ll}
\dfrac{(1-n)\,\Phi^2}{a^2\,\Psi^2}\Big[\dfrac{\Phi''}{\Phi}+2\,n\,\Big(\dfrac{\Phi'^2}{\Phi^2}\Big)-\dfrac{\Phi'\,\Psi'}{\Phi\,\Psi}\Big]
=\dfrac{\xi^2\,\Psi''}{\Psi}+n\,\Big(\dfrac{\xi^2\,\Phi''}{\Phi}\Big)-(n+1)\,\Big(\xi^2\,\dfrac{\Phi'\,\Psi'}{\Phi\,\Psi}\Big)\\
\\
\,\,\,\,\,\,\,\,\,\,
-2\,n\,\Big(\xi^2\,\dfrac{\Phi'^2}{\Phi^2}\Big)
+\dfrac{(1+6\,n)\,a-d}{a}\Big(\dfrac{\xi\,\Phi'}{\Phi}\Big)+\dfrac{(1+2\,n)\,a-d}{a}\Big(\dfrac{\Psi'}{\Psi}\Big)-6\,n-1+\dfrac{3\,d}{a},
\end{array}
\end{equation}
where $c=c_7$ and $d=c_8$ are arbitrary constants such that $f(r)=c\,\mathrm{exp}[d\,r]$. By integration the equation (\ref{u52-2}), we can get the following:
\begin{equation}\label{u54-2}
\begin{array}{ll}
\Psi(\xi)=q_1\,\Phi^{n-2}(\xi)\,\Phi'(\xi)\,\xi^{\tilde{d}\,\xi},
\end{array}
\end{equation}
where $\tilde{d}=2\,(2-n)+\dfrac{d\,(1-n)}{a\,n}$ while $q_1$ is an arbitrary constant of integration. Substitute (\ref{u54-2}) in (\ref{u53-2}), we have the following ordinary differential equation of the function $\Phi$ only as follows:
\begin{equation}\label{u55-2}
\begin{array}{ll}
a^2\,q_1^2\,\Phi^{2(n-3)}\,\xi^{1+2\,\tilde{d}}\,\Bigg[
2\,(1-n)\,(3\,n-4)\,\xi^2\,\Phi'^3-(n-1)\,\xi\,\Phi\,\Phi'\Big(
(1+5\,\tilde{d}-7\,n)\,\Phi'\\
\,\,\,\,\,\,\,\,\,\,\,\,\,\,\,
-(3\,n-7)\,\xi\,\Phi''\Big)+\Phi^2\Big(
\big[1-7\,n+\tilde{d}\,(5\,n-\tilde{d})\big]\,\Phi'+\xi\,\Big[
\big[(\tilde{d}+3)\,n-1-2\,\tilde{d}\big]\,\Phi''\\
\,\,\,\,\,\,\,\,\,\,\,\,\,\,\,
+(n-1)\,\xi\,\Phi'''\Big]
\Big)
\Bigg]=(1-n)^2\,\Big(\tilde{d}\,\Phi-(n+2)\,\xi\,\Phi'\Big),
\end{array}
\end{equation}
If one solves the above third order non-linear ordinary differential equation, he can obtain the exact solutions of the original Einstein field equations (\ref{u210})-(\ref{u213}) corresponding to reduction (\ref{u51-2}). This equation is very difficult to solve in general form, however, we can solve it in some special cases as follows:\\

Here, we assume the solution equation (\ref{u55-2}) in the form:
\begin{equation}\label{u56-2}
\begin{array}{ll}
\Phi(\xi)=q_2\,\xi^{q_3},
\end{array}
\end{equation}
where $q_2$ and $q_3$ are arbitrary non-zero constants. If we substitute the above solution in the equations (\ref{u55-2}), we have the following condition:
\begin{equation}\label{u57-2}
\begin{array}{ll}
2\,(n-1)^2\,q_2^{2\,(2-n)}\,(4\,n\,q_3+\alpha_1)\,\xi^{\alpha_1}=a_1^2\,q_1^2\,q_3(2\,q_3-\alpha_1-4)\,\big[4\,n\,(q_3-2)-\alpha_1\big],
\end{array}
\end{equation}
where $\alpha_1=2\big[(2-n)\,q_3-\tilde{d}\big]$. For discuss the above condition, we must take two cases as the following:\\

\textbf{Case (2.1):} When $\alpha_1\,\neq\,0$, then the condition (\ref{u57-2}) leads to two the following two conditions:
\begin{equation}\label{u58-2}
\begin{array}{ll}
4\,n\,q_3+\alpha_1=0,\,\,\,\,\,\,\,\,\,\,(2\,q_3-\alpha_1-4)\,\big[4\,n\,(q_3-2)-\alpha_1\big]=0.
\end{array}
\end{equation}
The solution of the above equation leads to the following two cases:\\

\textbf{Case (2.1.1):} The first solution is $q_3=1$ and $\alpha_1=-4\,n$ and the coefficients metric functions takes the form:
\begin{equation}\label{u59-2}
\left\{
  \begin{array}{ll}
A(r,t)=a_1\,(t+b)^{2\,n}\,\exp[a\,(1-2\,n)\,r],\\
\\
B(r,t)=a_2\,(t+b)\,\exp[a\,r],\\
\\
C(r,t)=a_3\,(t+b)^{n}\,\exp\Big[\dfrac{a\,n\,(1-2\,n)\,r}{n-1}\Big],
  \end{array}
\right.
\end{equation}
where $a_1=q_1\,q_2^{n-1}$, $a_2=q_2$ and $a_3=c_1\,q_2^{2\,n}$.  It is observed from the above equations, the line element (\ref{u21}) can be written in the following form:
\begin{equation}  \label{s21}
\begin{array}{ll}
ds_{21}^2=a_2^2\,(t+b)^2\,\exp[a\,r]\,dr^2-a_1^2\,(t+b)^{4\,n}\,\exp[2\,a\,(1-2\,n)\,r]\,dt^2\\
\\
\,\,\,\,\,\,\,\,\,\,\,\,\,\,\,\,\,\,\,\,\,\,\,\,\,\,\,\,\,\,\,\,\,\,\,\,\,\,\,\,\,\,\,\,\,\,\,\,\,\,
\,\,\,\,\,\,\,\,\,\,\,\,\,\,\,
+a_3^2\,(t+b)^{2\,n}\,\exp\Big[\dfrac{2\,a\,n\,(1-2\,n)\,r}{n-1}\Big]\,\Big(dx^2+dy^2\Big),
\end{array}
\end{equation}
where $a_1$, $a_2$, $a_3$, $a$ $b$ and $n$ are an arbitrary constants.\\

\textbf{Case (2.1.2):} The second solution of the equation (\ref{u58-2}) is $q_3=\dfrac{2}{2\,n+1}$ and $\alpha_1=-\dfrac{8\,n}{2\,n+1}$ and the coefficients metric functions $A(r,t)$ and $C(r,t)$ are functions of $t$ only, which case is not considered.\\

\textbf{Case (2.2):} When $\alpha_1\,=\,0$, then the condition (\ref{u57-2}) leads to $q_1=\pm\dfrac{(1-n)\,q_2^{2-n}}{a\,(q_3-2)}$. The coefficients metric functions takes the form:
\begin{equation}\label{u59-2-1}
\left\{
  \begin{array}{ll}
A(r,t)=a_1\,(t+b)^{q_3-1}\,\exp[\tilde{a}\,r],\\
\\
B(r,t)=a_2\,(t+b)^{q_3}\,\exp[\tilde{a}\,r],\\
\\
C(r,t)=a_3\,(t+b)^{n\,q_3}\,\exp\Big[\dfrac{\tilde{a}\,n\,r}{n-1}\Big],
  \end{array}
\right.
\end{equation}
where $\tilde{a}=a\,(q_3-2)$, $a_1=\dfrac{(1-n)\,q_2\,q_3}{\tilde{a}}$, $a_2=q_2$ and $a_3=c_1\,q_2^{2\,n}$.  It is observed from the above equations, the line element (\ref{u21}) can be written in the following form:
\begin{equation}  \label{s22}
\begin{array}{ll}
ds_{22}^2=a_2^2\,(t+b)^{2\,q_3}\,\exp[2\,\tilde{a}\,r]\Big[dr^2-\dfrac{(1-n)^2\,q_3^2}{\tilde{a}^2\,(t+b)^2}\,dt^2\Big]\\
\\
\,\,\,\,\,\,\,\,\,\,\,\,\,\,\,\,\,\,\,\,\,\,\,\,\,\,\,\,\,\,\,\,\,\,\,\,\,\,\,\,
\,\,\,\,\,\,\,\,\,\,\,\,\,\,\,\,\,\,\,\,\,\,\,\,\,\,\,\,\,\,\,\,\,
+a_3^2\,(t+b)^{2\,n\,q_3}\,\exp\Big[\dfrac{2\,\tilde{a}\,n\,r}{n-1}\Big]\,\Big(dx^2+dy^2\Big),
\end{array}
\end{equation}
where $a_2$, $a_3$, $q_3$, $\tilde{a}$, $b$ and $n$ are an arbitrary constants.\\

\textbf{Case (3):} When $c_3=0$ and $c_1\neq0$, the similarity variable and similarity functions can be written as the following:
\begin{equation}\label{u51-3}
\begin{array}{ll}
\xi=(r+a)\,\exp[b\,t],\,\,\,\,\,A(r,t)=\Psi(\xi)\,\exp[-b\,t],\,\,\,\,\,B(r,t)=\Phi(\xi),
\end{array}
\end{equation}
where $a=\dfrac{c_2}{c_1}$ and $b=-\dfrac{c_1}{c_4}$ are an arbitrary constants. Substituting the transformations (\ref{u51-3}) in the field Eqs. (\ref{u210-1}) and (\ref{u211-1}) lead
to the following system of ordinary differential equations:
\begin{equation}\label{u52-3}
\begin{array}{ll}
\dfrac{\xi\,\Phi''}{\Phi}-\dfrac{\xi\,\Phi'\,\Psi'}{\Phi\,\Psi}+(n-2)\,\dfrac{\xi\,\Phi'^2}{\Phi^2}
=\Big(\dfrac{d}{n}-d-1\Big)\,\dfrac{\Phi'}{\Phi},
\end{array}
\end{equation}

\begin{equation}\label{u53-3}
\begin{array}{ll}
\dfrac{(1-n)\,b^2\,\xi\,\Phi^2}{\Psi^2}\Big[\dfrac{\xi\,\Phi''+2\,\Phi'}{\Phi}+2\,n\,\xi\,\Big(\dfrac{\Phi'^2}{\Phi^2}\Big)-\dfrac{\xi\,\Phi'\,\Psi'}{\Phi\,\Psi}\Big]
=\dfrac{\Psi''}{\Psi}+n\,\Big(\dfrac{\Phi''}{\Phi}\Big)\\
\\
\,\,\,\,\,\,\,\,\,\,\,\,\,\,\,\,\,\,\,\,\,\,\,\,\,\,\,\,\,\,\,\,\,\,\,\,\,\,\,\,\,\,\,\,\,\,\,\,\,\,
-(n+1)\,\Big(\dfrac{\Phi'\,\Psi'}{\Phi\,\Psi}\Big)-2\,n\,\Big(\dfrac{\Phi'^2}{\Phi^2}\Big)
+\dfrac{d}{\xi}\,\Big(\dfrac{\Phi'}{\Phi}+\dfrac{\Psi'}{\Psi}+\dfrac{1}{\xi}\Big).
\end{array}
\end{equation}
where $f(r)=c\,(r+a)^d$ and $c=c_5\,c_1^{c_6}$ and $d=c_6$ are arbitrary constants. By integration the equation (\ref{u52-3}), we can get the following:
\begin{equation}\label{u54-3}
\begin{array}{ll}
\Psi(\xi)=q_1\,\xi^{\tilde{d}}\,\Phi^{n-2}(\xi)\,\Phi'(\xi),
\end{array}
\end{equation}
where $\tilde{d}=1+d+d/n$ while $q_1$ is an arbitrary constant of integration. Substitute (\ref{u54-3}) in (\ref{u53-3}), we have the following ordinary differential equation of the function $\Phi$ only as follows:

\begin{equation}\label{u55-3}
\begin{array}{ll}
q_1^2\,\Phi^{2(n-3)}\,\xi^{2\,\tilde{d}-3}\,\Bigg[
2\,(1-n)\,(3\,n-4)\,\xi^2\,\Phi'^3+(n-1)\,\xi\,\Phi\,\Phi'\Big(
(n-5\,\tilde{d})\,\Phi'\\
\,\,\,\,\,\,\,\,\,\,\,\,\,\,\,
+(3\,n-7)\,\xi\,\Phi''\Big)+\Phi^2\Big(
(1-\tilde{d})\,(n+\tilde{d})\,\Phi'+\xi\Big[\big[d\,(n-2)+n\big]\,\Phi''\\
\,\,\,\,\,\,\,\,\,\,\,\,\,\,\,
+(n-1)\,\xi\,\Phi'''\Big]
\Big)
\Bigg]=(1-n)^2\,b^2\Big(
(2-\tilde{d})\,\Phi-(n+2)\,\xi\,\Phi'\Big).
\end{array}
\end{equation}
It is important to note here that one can not solve eq. (56) in general.
So, in order to solve the problem completely, we have to choose some special cases as follows:

Here, we assume the solution equation (\ref{u55-3}) in the form:
\begin{equation}\label{u56-3}
\begin{array}{ll}
\Phi(\xi)=q_2\,\xi^{q_3},
\end{array}
\end{equation}
where $q_2$ and $q_3$ are arbitrary non-zero constants. If we substitute the above solution in the equations (\ref{u55-3}), we have the following condition:
\begin{equation}\label{u57-3}
\begin{array}{ll}
2\,b^2\,(4\,n\,q_3+\alpha_1)\,\xi^{\alpha_1}=q_1^2\,q_2^{2\,(n-2)}\,q_3\big[2\,(1+q_3)-\alpha_1\big]\,\big[4\,n\,(1+q_3)-\alpha_1\big],
\end{array}
\end{equation}
where $\alpha_1=2\,q_3\,(2+n)+4-2\,\tilde{d}$. For discuss the above condition, we must take two cases as the following:\\

\textbf{Case (3.1):} When $\alpha_1\,\neq\,0$, then the condition (\ref{u57-3}) leads to two the following conditions:
\begin{equation}\label{u58-3}
\begin{array}{ll}
4\,n\,q_3+\alpha_1=0,\,\,\,\,\,\,\,\,\,\,\big[2\,(1+q_3)-\alpha_1\big]\,\big[4\,n\,(1+q_3)-\alpha_1\big]=0.
\end{array}
\end{equation}
The solution of the above equations lead to the following solution:

\textbf{Case (3.1.1):} The first solution is $q_3=-\dfrac{1}{2}$ and $\alpha_1=2\,n$. Therefore, the coefficients metric functions takes the form:
\begin{equation}\label{u59-3}
\left\{
  \begin{array}{ll}
A(r,t)=a_1\,(r+a)^{\dfrac{1}{2}-n}\,\exp[\tilde{b}\,(2\,n+1)\,t],\\
\\
B(r,t)=a_2\,(r+a)^{-1/2}\,\exp[\tilde{b}\,t],\\
\\
C(r,t)=a_3\,(t+b)^{\dfrac{n\,(1-2\,n)}{2\,(n-1)}}\,\exp\Big[\tilde{b}\,n\,\,t\Big],
  \end{array}
\right.
\end{equation}
where $\tilde{b}=-\dfrac{b}{2}$, $a_1=-\dfrac{1}{2}\,q_1\,q_2^{n-1}$, $a_2=q_2$ and $a_3=c_1\,q_2^{n}$.  It is observed from the above equations, the line element (\ref{u21}) can be written in the following form:
\begin{equation}  \label{s31}
\begin{array}{ll}
ds_{31}^2=\dfrac{a_2^2}{r+a}\,\exp[2\,\tilde{b}\,t]\,dr^2-a_1^2\,(r+a)^{1-2\,n}\,\exp[2\,\tilde{b}\,(2\,n+1)\,t]\,dt^2\\
\\
\,\,\,\,\,\,\,\,\,\,\,\,\,\,\,\,\,\,\,\,\,\,\,\,\,\,\,\,\,\,\,\,\,\,\,\,\,\,\,\,\,\,\,\,
\,\,\,\,\,\,\,\,\,\,\,\,\,\,\,\,\,\,\,\,\,\,\,\,\,\,\,\,\,\,\,\,
+a_3^2\,(r+a)^{\dfrac{n\,(1-2\,n)}{(n-1)}}\,\exp\Big[2\,\tilde{b}\,n\,\,t\Big]\Big(dx^2+dy^2\Big),
\end{array}
\end{equation}
where $a_1$, $a_2$, $a_3$, $a$ $b$ and $n$ are an arbitrary constants.\\

\textbf{Case (3.1.1):} The second solution of equation (\ref{u58-3}) is $q_3=-\dfrac{1}{2\,n+1}$ and $\alpha_1=\dfrac{4\,n}{2\,n+1}$. In this case, the function $\Psi(\xi)$ is a constant which is not considered.\\

\textbf{Case (3.2):} When $\alpha_1=0$, the solution of the equation (\ref{u57-3}) is $q_1=\pm\,\dfrac{b\,(n-1)\,q_2^{2-n}}{q_3+1}$. Therefore, the coefficients metric functions takes the form:
\begin{equation}\label{u511-3}
\left\{
  \begin{array}{ll}
A(r,t)=a_1\,(r+a)^{q_3+1}\,\exp\big[\tilde{b}\,t\big],\\
\\
B(r,t)=a_2\,(r+a)^{q_3}\,\exp\big[\tilde{b}\,t\big],\\
\\
C(r,t)=a_3\,(t+b)^{\dfrac{n\,(1+q_3)}{n-1}}\,\exp\big[n\,\tilde{b}\,t\big],
  \end{array}
\right.
\end{equation}
where $\tilde{b}=q_3\,b$, $a_1=\dfrac{(n-1)\,q_2\,\tilde{b}}{q_3+1}$, $a_2=q_2$ and $a_3=c_1\,q_2^n$.  It is observed from the above equations, the line element (\ref{u21}) can be written in the following form:
\begin{equation}  \label{s32}
\begin{array}{ll}
ds_{32}^2=a_2^2\,(r+a)^{2\,q_3}\,\exp\big[2\,\tilde{b}\,t\big]\Big[dr^2-\Big(\dfrac{(n-1)\,\tilde{b}}{q_3+1}\Big)^2\,dt^2\Big]\\
\\
\,\,\,\,\,\,\,\,\,\,\,\,\,\,\,\,\,\,\,\,\,\,\,\,\,\,\,\,\,\,
\,\,\,\,\,\,\,\,\,\,\,\,\,\,\,\,\,\,\,\,\,\,\,\,\,\,\,\,\,\,\,\,\,\,\,\,\,\,\,\,
+a_3^2\,(t+b)^{\dfrac{2\,n\,(1+q_3)}{n-1}}\,\exp\big[2\,n\,\tilde{b}\,t\big]\Big(dx^2+dy^2\Big),
\end{array}
\end{equation}
where $a_2$, $a_3$, $a$ $b$, $q_3$ and $n$ are an arbitrary constants.\\

\textbf{Case (4):} When $c_1\neq0$ and $c_3\neq0$, the similarity variable and similarity functions can be written as the following:
\begin{equation}\label{u51-4}
\begin{array}{ll}
\xi=(t+b)\,(r+a)^c,\,\,\,\,\,A(r,t)=\Psi(\xi)\,(r+a)^{1-c},\,\,\,\,\,B(r,t)=\Phi(\xi)\,(r+a)^{-2\,c},
\end{array}
\end{equation}
where $a=\dfrac{c_2}{c_1}$, $b=\dfrac{c_4}{c_3}$ and $c=-\dfrac{c_3}{c_1}$ are an arbitrary constants. Substituting the transformations (\ref{u51-4}) in the field eqs. (\ref{u210-1}) and (\ref{u211-1}) lead
to the following system of ordinary differential equations:
\begin{equation}\label{u52-4}
\begin{array}{ll}
\dfrac{\Phi''}{\Phi}-\dfrac{\Phi'\,\Psi'}{\Phi\,\Psi}+(n-2)\,\dfrac{\Phi'^2}{\Phi^2}
=\Big(\dfrac{k\,(n-1)-n\,\big[1+2\,c\,(n-2)\big]}{n\,c\,\xi}\Big)\,\dfrac{\Phi'}{\Phi},
\end{array}
\end{equation}

\begin{equation}\label{u53-4}
\begin{array}{ll}
\dfrac{(1-n)\,\Phi^2}{c^2\,\Psi^2}\Big[\dfrac{\Phi''}{\Phi}+2\,n\,\Big(\dfrac{\Phi'^2}{\Phi^2}\Big)-\dfrac{\Phi'\,\Psi'}{\Phi\,\Psi}\Big]
=\xi^2\Bigg[\dfrac{\Psi''}{\Psi}+n\,\Big(\dfrac{\Phi''}{\Phi}\Big)-(n+1)\,\Big(\dfrac{\Phi'\,\Psi'}{\Phi\,\Psi}\Big)\\
\\
\,\,\,\,\,\,\,\,\,\,\,\,\,\,\,
-2\,n\,\Big(\dfrac{\Phi'^2}{\Phi^2}\Big)\Bigg]
+\xi\Bigg[\dfrac{c-1-k+2\,n\,(3\,c-1)}{c}\,\Big(\dfrac{\Phi'}{\Phi}+\dfrac{1-k+c\,(1+2\,n)}{c}\Big(\dfrac{\Psi'}{\Psi}\Big)\Bigg]\\
\\
\,\,\,\,\,\,\,\,\,\,\,\,\,\,\,
-\dfrac{2\,k-c\,(1+3\,k+4\,n)+c^2\,(6\,n+1)}{c^2}.
\end{array}
\end{equation}
where $f(r)=d\,(r+a)^k$ and $d=c_5\,c_1^{c_6}$ and $k=c_6$ are arbitrary constants. By integration the equation (\ref{u52-4}), we can get the following:
\begin{equation}\label{u54-4}
\begin{array}{ll}
\Psi(\xi)=q_1\,\xi^{\tilde{k}}\,\Phi^{n-2}(\xi)\,\Phi'(\xi),
\end{array}
\end{equation}
where $\tilde{k}=2\,(2-n)+\dfrac{k-1}{c}-\dfrac{k}{n\,c}$ while $q_1$ is an arbitrary constant of integration. Substitute (\ref{u54-4}) in (\ref{u53-4}), we have the following ordinary differential equation of the function $\Phi$ only as follows:

\begin{equation}\label{u55-4}
\begin{array}{ll}
q_1^2\,\Phi^{2\,(n-3)}\,\xi^{2\,\tilde{k}+1}\,\Bigg[
2\,c^2\,(n-1)\,(3\,n-4)\,\xi^2\,\Phi'^3
+c\,(n-1)\,\xi\,\Phi\,\Phi'\Big(\big[
5+c\,(1+5\,\tilde{k})\\
\,\,\,\,\,\,\,\,\,\,+n\,(2-7\,c)\big]\Phi'
+c\,(7-3\,n)\,\xi\,\Phi''\Big)
+\Phi^2\Big(
\Big[2\,n+\big(1+\tilde{k}+2\,n\,(\tilde{k}-4)\\
\,\,\,\,\,\,\,\,\,\,+c\,\big[\tilde{k}^2-1+n\,(7-5\,\tilde{k})\big]\big)\Big]\,\Phi'
+c\Big[\big[1+c\,(1+2\,\tilde{k})-c\,n\,(3+\tilde{k})\big]\,\Phi''\\
\,\,\,\,\,\,\,\,\,\,
-c\,(n-1)\,\xi\,\Phi'''\Big]\Big)\Bigg]=(n-1)^2\big[(2+n)\,\xi\,\Phi'-\tilde{k}\,\Phi\big].
\end{array}
\end{equation}
If one solves the above third order non-linear ordinary differential equation, he can obtain the exact solutions of the original Einstein field equations (\ref{u210})-(\ref{u213}) corresponding to reduction (\ref{u51-4}). This equation is very difficult to solve in general form, however, we can solve it in some special cases as follows:\\

Here, we assume the solution equation (\ref{u55-4}) in the form:
\begin{equation}\label{u56-4}
\begin{array}{ll}
\Phi(\xi)=q_2\,\xi^{q_3},
\end{array}
\end{equation}
where $q_2$ and $q_3$ are arbitrary non-zero constants. If we substitute the above solution in the equations (\ref{u55-4}), we have the following condition:
\begin{equation}\label{u57-4}
\begin{array}{ll}
q_1^2\,q_2^{2\,(n-2)}\,q_3\,\Big[4\,n\,\big[1+c\,(q_3-2)\big]+c\,\alpha_1\Big]\,\big[2+c\,(2\,q_3+\alpha_1-4)\big]\,\xi^{\alpha_1}\\
\,\,\,\,\,\,\,\,\,\,\,\,\,\,\,\,\,\,\,\,\,\,\,\,\,\,\,\,\,\,\,\,\,\,\,\,\,\,\,\,\,\,\,\,\,\,\,\,\,\,\,\,\,\,\,\,\,\,\,\,
=2\,(1-n)^2\,(4\,n\,q_3-\alpha_1),
\end{array}
\end{equation}
where $\alpha_1=2\,\big[\tilde{k}+q_3\,(n-2)\big]$. For discuss the above condition, we must take two cases as the following:\\

\textbf{Case (4.1):} When $\alpha_1\neq0$, then the condition (\ref{u57-4}) leads to two the following conditions:
\begin{equation}\label{u58-4}
\begin{array}{ll}
\Big[4\,n\,\big[1+c\,(q_3-2)\big]+c\,\alpha_1\Big]\,\big[2+c\,(2\,q_3+\alpha_1-4)\big]=0,\,\,\,\,\,\,\,4\,n\,q_3-\alpha_1=0.
\end{array}
\end{equation}
The above conditions leads to the following two solutions:\\

\textbf{Case (4.1.1):} The first solution is $q_3=1-\dfrac{1}{2\,c}$ and $\alpha_1=\dfrac{(2\,n\,(2\,c-1)}{c}$. Therefore, the coefficients metric functions takes the form:
\begin{equation}\label{u510-4}
\left\{
  \begin{array}{ll}
A(r,t)=a_1\,(r+a)^{m_1}\,(t+b)^{m\,(1+2\,n)-1},\\
\\
B(r,t)=a_2\,(r+a)^{m_2}\,(t+b)^{m},\\
\\
C(r,t)=a_3\,(r+a)^{m_3}\,(t+b)^{n\,m},
  \end{array}
\right.
\end{equation}
where $c=\dfrac{1}{2\,(1-m)}$, $a_1=m\,q_1\,q_2^{n-1}$, $a_2=q_2$, $a_3=d\,q_2^{n}$, $m_1=\dfrac{m\,(1-2\,n)}{2\,(1-m)}$, $m_2=\dfrac{2-m}{2\,(m-1)}$ and $m_3=\dfrac{n\,m_1}{(n-1)}$.   It is observed from the above equations, the line element (\ref{u21}) can be written in the following form:
\begin{equation}  \label{s41}
\begin{array}{ll}
ds_{41}^2=a_2^2\,(r+a)^{2\,m_2}\,(t+b)^{2\,m}\,dr^2-a_1^2\,(r+a)^{2\,m_1}\,(t+b)^{2\,m\,(1+2\,n)-2}\,dt^2\\
\,\,\,\,\,\,\,\,\,\,\,\,\,\,\,\,\,\,\,\,\,\,\,\,\,\,\,\,\,\,\,\,\,\,\,\,\,\,\,\,
\,\,\,\,\,\,\,\,\,\,\,\,\,\,\,\,\,\,\,\,\,\,\,\,\,\,\,\,\,\,\,\,\,\,\,\,\,\,\,\,\,\,\,
+a_3^2\,(r+a)^{2\,m_3}\,(t+b)^{2\,n\,m}\,\Big(dx^2+dy^2\Big),
\end{array}
\end{equation}
where $a_1$, $a_2$, $a_3$, $a$, $b$, $m$ and $n$ are an arbitrary constants.\\

\textbf{Case (4.1.2):} The second solution of the equation (\ref{u58-4}) is $q_3=\dfrac{2\,c-1}{c\,(2\,n+1)}$ and $\alpha_1=\dfrac{4\,n\,(2\,c-1)}{c\,(1+2\,n)}$. In this case the coefficients metric functions $A$ and $C$ are function of $t$ only, and then not considered here.\\

\textbf{Case (4.2):} When $\alpha_1=0$, then $q_1=\pm\dfrac{(n-1)\,q_2^{2-n}}{1+c\,(q_3-2)}$. Therefore, the coefficients metric functions takes the form:
\begin{equation}\label{u511-4-1}
\left\{
  \begin{array}{ll}
A(r,t)=a_1\,(r+a)^{m_1}\,(t+b)^{m_2},\\
\\
B(r,t)=a_2\,(r+a)^{m_1-1}\,(t+b)^{m_2+1},\\
\\
C(r,t)=a_3\,(r+a)^{\dfrac{n\,m_1}{n-1}}\,(t+b)^{n\,(1+m_2)},
  \end{array}
\right.
\end{equation}
where $m_1=1+c\,(q_3-2)$, $m_2=q_3-1$, $a_1=\dfrac{(n-1)\,(1+m_2)\,q_2}{m_1}$, $a_2=q_2$ and $a_3=d_1\,q_2^n$.  It is observed from the above equations, the line element (\ref{u21}) can be written in the following form:
\begin{equation}  \label{s42}
\begin{array}{ll}
ds_{42}^2=a_2^2\,(r+a)^{2\,(m_1-1}\,(t+b)^{2\,m_2}\,\Big[(t+b)^2\,dr^2-\dfrac{(n-1)^2\,(1+m_2)^2}{m_1^2}\,dt^2\Big]\\
\,\,\,\,\,\,\,\,\,\,\,\,\,\,\,\,\,\,\,\,\,\,\,\,\,\,\,\,\,\,\,\,\,\,\,\,\,\,\,\,
\,\,\,\,\,\,\,\,\,\,\,\,\,\,\,\,\,\,\,\,\,\,\,\,\,\,\,\,\,\,\,\,\,\,\,\,\,\,\,\,
+a_3^2\,(r+a)^{\dfrac{2\,n\,m_1}{n-1}}\,(t+b)^{2\,n\,(1+m_2)}\,\Big(dx^2+dy^2\Big),
\end{array}
\end{equation}
where $a_2$, $a_3$, $a$, $b$, $m_1$, $m_2$ and $n$ are an arbitrary constants.\\

\section{Some physical and geometric features:}
In this section, we study some parameters of the models obtained in the proceeding sections.

\subsection{Parameters of the model (\ref{s21}):}

For the line element (\ref{s21}), the expressions for density and pressure are given by:
\begin{equation}  \label{u61-2}
  \begin{array}{ll}
\chi\,\rho(r,t)\,=\dfrac{n\,(n+2)}{a_1^2\,(t+b)^{2}}\,\Bigg[(t+b)^{-4\,n}\,\exp[4\,a\,n\,r]+K_1\Bigg]\,\exp[-2\,a\,r],\\
\\
\chi\,p(r,t)\,=\dfrac{n\,(n+2)}{a_1^2\,(t+b)^{2}}\,\Bigg[(t+b)^{-4\,n}\,\exp[4\,a\,n\,r]+\Big(\dfrac{1-2\,n}{1+2\,n}\Big)\,K_1\Bigg]\,\exp[-2\,a\,r],
\end{array}
\end{equation}
where $K_1=\dfrac{(1-2\,n)\,(1+2\,n)\,(3\,n-2)\,a^2\,a_1^2}{(n+2)(n-1)^2\,a_2^2}$.

The volume element is
\begin{equation}  \label{u62-2}
V=a_1\,a_2\,a_3^2\,(t+b)^{4\,n+1}\,\exp\Bigg[2\,a\,\Big(\dfrac{1+3\,n\,(n-1)}{n-1}\Big)\,r\Bigg].
\end{equation}
The expansion scalar, which determines the volume behavior of the fluid, is given by:
\begin{equation}\label{u63-2}
  \begin{array}{ll}
\Theta=\dfrac{2\,n+1}{a_1\,(t+b)^{2\,n+1}}\,\exp\big[a\,(2\,n-1)\,r\big].
  \end{array}
\end{equation}

\subsection{Parameters of the model (\ref{s31}):}

For the line element (\ref{s31}), the expressions for density and pressure are given by:
\begin{equation}  \label{u61-3}
  \begin{array}{ll}
\chi\,\rho(r,t)\,=\dfrac{n\,(n+2)\,\tilde{b}^2}{a_1^2\,(r+a)}\,\Bigg[(r+a)^{2\,n}\exp[-4\,\tilde{b}\,t]+K_1\Bigg]\,\exp[-2\,\tilde{b}\,t],\\
\\
\chi\,p(r,t)\,=\dfrac{n\,(n+2)\,\tilde{b}^2}{a_1^2\,(r+a)}\,\Bigg[(r+a)^{2\,n}\exp[-4\,\tilde{b}\,t]
+\Big(\dfrac{1-2\,n}{1+2\,n}\Big)\,K_1\Bigg]\,\exp[-2\,\tilde{b}\,t],
\end{array}
\end{equation}
where $K_1=\dfrac{(1-2\,n)\,(1+2\,n)\,(3\,n-2)\,a_1^2}{4\,(n+2)\,(n-1)^2\,a_2^2\,\tilde{b}^2}$.

The volume element is
\begin{equation}  \label{u62-3}
V=a_1\,a_2\,a_3^2\,(r+a)^{\dfrac{n\,(2-3\,n)}{n-1}}\,\exp\Big[2\,(2\,n+1)\,\tilde{b}\,t\Big].
\end{equation}
The expansion scalar, which determines the volume behavior of the fluid, is given by:
\begin{equation}\label{u63-3}
  \begin{array}{ll}
\Theta=\dfrac{(2\,n+1)\,\tilde{b}}{a_1}\,(r+a)^{\dfrac{2n-1}{2}}\,\exp\big[-\tilde{b}\,(2\,n+1)\,t\big].
  \end{array}
\end{equation}

\subsection{Parameters of the model (\ref{s41}):}

For the line element (\ref{s41}), the expressions for density and pressure are given by:
\begin{equation}  \label{u61-4}
  \begin{array}{ll}
\chi\,\rho(r,t)\,=\dfrac{n\,m\,(n+2)}{a_1^2}\,\Bigg[(r+a)^{M_1}\,(t+b)^{-4\,n\,m}+K_1\Bigg]\,(r+a)^{M_2}\,(t+b)^{-2\,m},\\
\\
\chi\,p(r,t)\,=\dfrac{n\,m^2\,(n+2)}{a_1^2}\,\Bigg[(r+a)^{M_1}\,(t+b)^{-4\,n\,m}
+\Big(\dfrac{1-2\,n}{1+2\,n}\Big)\,K_1\Bigg]\,(r+a)^{M_2}\,(t+b)^{-2\,m},
\end{array}
\end{equation}
where $K_1=\dfrac{(1-2\,n)\,(1+2\,n)\,(3\,n-2)\,a_1^2}{(n+2)\,(n-1)^2\,(m-1)^2\,a_2^2}$, $M_1=\dfrac{2\,n\,m}{m-1}$ and $M_2=\dfrac{m}{1-m}$.

The volume element is
\begin{equation}  \label{u62-4}
V=a_1\,a_2\,a_3^2\,(r+a)^{n\,m\,(2-3\,n)+n-1}\,(t+b)^{2\,m\,(1+2\,n)-1}.
\end{equation}
The expansion scalar, which determines the volume behavior of the fluid, is given by:
\begin{equation}\label{u63-4}
  \begin{array}{ll}
\Theta=\dfrac{m\,(2\,n+1)}{a_1}\,(r+a)^{\dfrac{m\,(1-2\,n)}{2\,(1-m)}}\,(t+b)^{-m\,(1+2\,n)}.
  \end{array}
\end{equation}

\textbf{Remark (1):} For the models (\ref{s1}), (\ref{s22}), (\ref{s32}) and (\ref{s42}), the density and 
the pressure is vanishes identically and hence these are the vacuum plane symmetric models of universe.\\

\textbf{Remark (2):} For the models (\ref{s21}), (\ref{s31}) and (\ref{s41}), the non-vanishing components of the shear tensor, $\sigma_i^j$, are:
\begin{equation}\label{u64-2}
  \begin{array}{ll}
\sigma_3^3\,=\,\sigma_2^2\,=\,-\dfrac{1}{2}\,\sigma_1^1\,=\,\dfrac{(n-1)\,\Theta}{3\,(2\,n+1)}.
  \end{array}
\end{equation}
The shear scalar $\sigma$, is given by:
\begin{equation}\label{u65-2}
  \begin{array}{ll}
\sigma^2\,=\dfrac{(n-1)^2\,\Theta^2}{3\,(2\,n+1)^2}.
  \end{array}
\end{equation}
The deceleration parameter is given by Feinstien and Ibanez \cite{fein1}:
\begin{equation}\label{u66-2}
  \begin{array}{ll}
\mathbf{q}\,=-3\,\Theta^2\,\Big(\Theta_{;i}\,u^i+\dfrac{1}{3}\,\Theta^2\Big)
=2\,\Theta^4.
  \end{array}
\end{equation}

\textbf{Remark (3):} All models do not admit rotation, since $\omega_{ij}=0$.

\section{Conclusion}

We have investigated a plane symmetric inhomogeneous cosmological models with a perfect fluid in general relativity 
which are based on exact solution of Einstein's equations. 
In these models, the matter source consists of perfect fluid with proportionality relation between expansion scalar
and shear scalar. The Lie group analysis transform the system of partial differential equations (Einstein field equations) 
to the system of ordinary differential equations. By choosing some special cases, 
we obtained a new class of exact solutions for the Einstein field equations for 
these models using the symmetry group analysis method. 
For $n > 0$ and $m > 0$, the derived models have big bang singularity at $t = -b$.
Since, in our analysis, $n$ is taken as positive so the values of $m$ control the 
singularity nature of derived models. For $m >0$, all the models have singular origin 
while for $m \leq 0$, the models have non singular origin. In the literature, the non singular 
models seem reasonable to project the dynamics of future universe while the singular models seem to 
describe the dynamics of the universe from big bang to present epoch.\\


\end{document}